\documentclass[journal=jacsat,manuscript=article]{achemso}

\usepackage{chemformula} % Formula subscripts using \ch{}
\usepackage[T1]{fontenc} % Use modern font encodings
\usepackage{braket}
\usepackage[colorlinks,allcolors=blue]{hyperref}

\author{Aurélie Champagne}
\affiliation[LBNL]{Materials and Chemical Sciences Division, Lawrence Berkeley National Laboratory, Berkeley, CA 94720, USA}
\alsoaffiliation[UCBerkeley]{Department of Physics, University of California Berkeley, Berkeley, CA 94720, USA}
\email{achampagne@lbl.gov}
\author{Jonah B. Haber}
\affiliation[UCBerkeley]{Department of Physics, University of California Berkeley, Berkeley, CA 94720, USA}
\author{Supavit Pokawanvit}
\affiliation[Stanford1]{Department of Materials Science and Engineering, Stanford University, Stanford, CA 94305, USA}
\alsoaffiliation[Stanford2]{Department of Applied Physics, Stanford University, Stanford, CA 94305, USA}
\author{Diana Y. Qiu}
\affiliation[Yale1]{Department of Mechanical Engineering and Materials Science, Yale University, New Haven, CT 06520, USA}
\alsoaffiliation[Yale2]{Energy Science Institute, Yale University, New Haven, CT 06520, USA}
\author{Souvik Biswas}
\affiliation[Caltech]{Thomas J. Watson Laboratory of Applied Physics, Caltech, Pasadena, CA 91125, USA}
\author{Harry A. Atwater}
\affiliation[Caltech]{Thomas J. Watson Laboratory of Applied Physics, Caltech, Pasadena, CA 91125, USA}
\author{Felipe H. da Jornada}
\affiliation[Stanford]{Department of Materials Science and Engineering, Stanford University, Stanford, CA 94305, USA}
\author{Jeffrey B. Neaton}
\affiliation[LBNL]{Materials and Chemical Sciences Division, Lawrence Berkeley National Laboratory, Berkeley, CA 94720, USA}
\alsoaffiliation[UCBerkeley]{Department of Physics, University of California Berkeley, Berkeley, CA 94720, USA}
\alsoaffiliation[Kavli]{Kavli Energy Nanosciences Institute at Berkeley, Berkeley, CA 94720, USA}
\email{jbneaton@lbl.gov}

\title{Quasiparticle and Optical Properties of Carrier-Doped Monolayer MoTe$_2$ from First Principles}

\abbreviations{2D, BSE, DFT, PPM, QP}

\begin{document}

\begin{abstract}
The intrinsic weak and highly non-local dielectric screening of two-dimensional materials is well known to lead to high sensitivity of their optoelectronic properties to environment. Less studied theoretically is the role of free carriers on those properties. Here, we use \textit{ab initio} GW and Bethe-Salpeter equation calculations, with a rigorous treatment of dynamical screening and local-field effects, to study the doping-dependence of the quasiparticle and optical properties of a monolayer transition metal dichalcogenide, 2H MoTe$_2$. We predict a quasiparticle band gap renormalization of several hundreds meV for experimentally-achievable carrier densities, and a similarly sizable decrease in the exciton binding energy. This results in an almost constant excitation energy for the lowest-energy exciton resonance with increasing doping density. Using a newly-developed and generally-applicable quasi-2D plasmon-pole model and a self-consistent solution of the Bethe-Salpeter equation, we reveal the importance of accurately capturing both dynamical and local-field effects to understand detailed photoluminescence measurements.
\end{abstract}

In two-dimensional (2D) semiconducting materials, quantum confinement and weak and highly non-local dielectric screening have lead to unprecedented optoelectronic phenomena, including the existence of strongly bound electron-hole pairs, or excitons~\cite{keldysh1979,splendiani2010,mak2013,thygesen2017}, with binding energies that can reach several hundreds of meV in monolayer transition metal dichalcogenides (TMDs)~\cite{ramasubramaniam2012,qiu2013,mak2013,chernikov2014,ye2014,ugeda2014,he2014,wang2015}. While the band gap of bulk semiconductors is usually not strongly affected by extrinsic factors, the weak intrinsic screening of atomically thin materials renders the quasiparticle (QP) gap in 2D materials highly sensitive to their surrounding dielectric and electronic environment~\cite{mak2012,mak2013,ross2013,falco2013,ugeda2014,plechinger2015,shang2015}. In particular, the presence of charged free carriers (electrons or holes) can screen electron-hole interactions and modify the QP band gap, as well as optical properties, \textit{i.e.}, photoexcitation energies and oscillator strengths~\cite{ugeda2014,liang2015,gao2016}. Experimentally, monolayer TMDs can be unintentionally heavily carrier-doped which renders the interpretation of measurements of their electronic and optical spectra non-trivial~\cite{radisavljevic2011,radisavljevic2013,mak2013,mouri2013}.\ 

First-principles calculations can guide understanding of the role of screening effects on optoelectronic properties of low-dimensional systems. The \textit{ab initio} GW and GW plus Bethe-Salpeter equation (GW-BSE) approach is a state-of-the-art method for the accurate prediction of one and two particle excitations, respectively~\cite{hedin1965,strinati1982,hybertsen1986,rohlfing2000}. Central to this approach is the computation of the dielectric function which encodes how the many-electron system screens charged and neutral excitations. Plasmon-pole models~\cite{hybertsen1986,godby1989,oschlies1995} are commonly used to efficiently capture the dynamical features of the dielectric response of undoped materials. However, for the description of doped semiconductors, these models need to be modified to include the effects of additional low-energy acoustic carrier plasmon that strongly couples with QP and neutral excitations. Moreover, while standard BSE calculations are typically performed in the static limit, where the exciton binding energy is much smaller than the plasma frequency, the static approximation is no longer adequate for semiconductors degenerately doped with a finite carrier density.\

\begin{figure*}[h!]
	\includegraphics[width=0.98\textwidth]{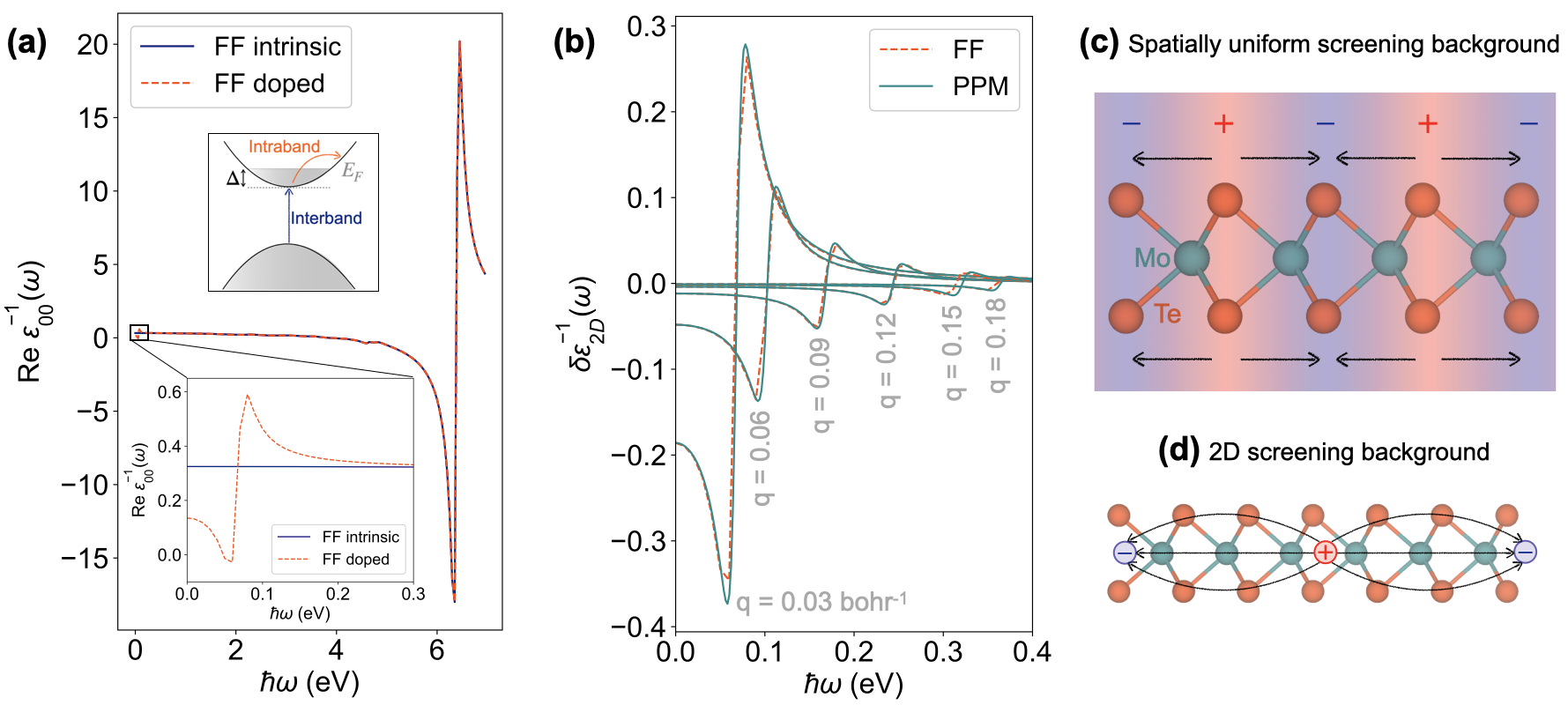}
	\centering
	\caption{(a) Real part of the head element ($\mathbf{G}=\mathbf{G}'=0$) of the inverse dielectric matrix, $\epsilon^{-1}_{00}(\omega)$, computed from an explicit, full frequency-dependent (FF) calculation for the intrinsic (solid blue curve) and doped MoTe$_2$ monolayer systems (dashed orange curve), highlighting the presence of a low-frequency carrier plasmon-pole arising from intraband transitions, well-separated from the high-frequency intrinsic plasmon-pole from interband transitions (inset). (b) Difference between the doped and intrinsic effective 2D inverse dielectric matrices, $\delta \epsilon_{\mathrm{2D}}^{-1}(\mathbf{q},\omega) = \epsilon_{\mathrm{2D}}^{-1}(\mathbf{q},\omega) - \epsilon_{\mathrm{int},\mathrm{2D}}^{-1}(\mathbf{q},\omega)$~\cite{qiu2016}, for a set of $\mathbf{q}$ wavevectors, obtained with our PPM (solid cyan curve), and compared to a FF calculation (dashed orange curve) for the MoTe$_2$ monolayer. Schematics of the electron-hole interaction with (c) a spatially uniform screening background (no local fields), and (d) a 2D screening background with non-negligible off-diagonal elements of the dielectric matrix (local-field effects).}
	\label{fig1}
\end{figure*}

Here, we detail the development of a general, accurate, and computationally-efficient approach for predicting excitonic effects in carrier-doped materials using the \textit{ab initio} GW-BSE method. We illustrate this approach for doped MoTe$_2$ monolayers, a less studied member of the monolayer TMD class, comparing to detailed spectroscopy measurements~\cite{biswas2023}. Central to our approach is a modified plasmon-pole model (PPM) which accurately captures two poles in the frequency-dependent dielectric function, \textit{i.e.}, the high-frequency optical intrinsic plasmon and the low-frequency acoustic carrier plasmon (Figure~\ref{fig1}). Our modified PPM only requires the explicit calculation of the dielectric function at two real frequencies, significantly reducing the computational time and memory usage with respect to an explicit, fully frequency-dependent (FF) calculation without any PPM. Our model reveals the importance of including dynamical effects and local-field effects in the effective screened electron-hole interaction, as well as a self-consistent solution to the BSE, to obtain a good agreement with experimental photoluminescence spectra. Our new PPM approach should be applicable for obtaining the optical absorption spectrum of a variety of doped 2D materials with delocalized excitons or van der Waals homobilayers.\

In the GW approximation, the self-energy operator, $\Sigma$, is expanded to first-order in the screened Coulomb interaction, $W$, and single-particle Green's function, $G$, explicitly $\Sigma = iGW$~\cite{hedin1965, strinati1982, hybertsen1986}. In reciprocal space and with a plane-wane basis, the screened Coulomb interaction is related to the inverse dielectric matrix $\epsilon^{-1}$ through
\begin{equation}
\label{eq1}
    W_{\mathbf{GG}'}(\mathbf{q},\omega) = \epsilon_{\mathbf{GG}'}^{-1}(\mathbf{q},\omega) v(\mathbf{q}+\mathbf{G}'),
\end{equation}
where $v(\mathbf{q}+\mathbf{G'}) = 4\pi/|\mathbf{q}+\mathbf{G}|^2$ is the bare Coulomb interaction, and $\mathbf{G}$ and $\mathbf{G}'$ are reciprocal lattice vectors. The inverse dielectric function, $\epsilon_{\mathbf{GG}'}^{-1}(\mathbf{q},\omega)$, is typically computed within the random phase approximation~\cite{hybertsen1987} which captures both dynamical ($\omega$) and local-field (off diagonal in $\mathbf{GG}'$) effects. While an explicit FF calculation would give an accurate QP band gap for doped semiconductors, in practice both the fine frequency sampling necessary to capture the carrier plasmon pole as well as the ultra-fine Brillouin zone $k$-point sampling required to cover the experimentally-accessible doping concentrations render such explicit calculations impractical. Low-rank approximations to dielectric function, like the one used in the static subspace approach, can speed-up full-frequency GW calculations $3-5$ times~\cite{nguyen2012,delben2019,shao2016}; however, even so, calculations still remain computationally expensive for doped systems. A practical alternative is to extend the static dielectric function to finite frequencies using a plasmon-pole model, such as, for example, Hybertsen-Louie~\cite{hybertsen1986} or Godby-Needs~\cite{godby1989,oschlies1995}. Such models are routinely used to study intrinsic semiconductors, and have been partially adapted for doped systems to capture the carrier plasmon-pole at low frequency (Figure~\ref{fig1})~\cite{ando1982,sarma1996,spataru2010,spataru2013,liang2015}. An important first step was taken by Gao \textit{et al.}~\cite{gao2016} to study the effects of carrier doping in monolayer MoS$_2$. In their approach, they include two important approximations: first, they use a single plasmon-pole model which effectively lumps together the intrinsic and carrier plasmons; and second, when solving the BSE, they consider a spatially uniform screening background approximation, shown schematically in Figure~\ref{fig1}c, where they assume the additional charge carriers modify only the head of the dielectric matrix, $\epsilon_{00}^{-1}(\mathbf{q},\omega)$. While such an approach has been successful to qualitatively and quantitatively describe the renormalization of the optical absorption spectrum of monolayer MoS$_2$ with carrier doping, the validity of these two critical approximations has not been assessed. Below we describe our new PPM that maintains two physically distinct poles for the intrinsic and carrier plasmon, and explicitly incorporates how carrier doping changes the entire dielectric matrix, resulting in greater accuracy and predictive power for the exciton energy as well as for the QP band gap and the exciton binding energy. In particular, we find that for monolayer MoTe$_2$ it is crucial to include local-field effects to describe the screening from the free carriers, otherwise the renormalization of the exciton binding energy may be underestimated by a factor from $2$ to $3$.\

Neutral excitations of 2D semiconductors can be obtained by solving the BSE for the two-particle Green's function in the electron-hole subspace. When performed atop a GW calculation with $G$ and $W$ constructed from density functional theory (DFT) calculations, the method is commonly referred to as the \textit{ab initio} GW-BSE approach~\cite{hedin1965,strinati1982,rohlfing2000}. Within the Tamm-Dancoff approximation, the BSE can be cast as a generalized eigenvalue problem in a basis of electron-hole product states. Explicitly, in reciprocal space
\begin{equation}
\label{eq5}
    \left( E_{c\mathbf{k}} - E_{v\mathbf{k}} \right) A_{vc\mathbf{k}}^S + \sum_{v'c'\mathbf{k}'} K_{vc\mathbf{k},v'c'\mathbf{k}'}(\Omega^S) A_{v'c'\mathbf{k}'}^S = \Omega^S A_{vc\mathbf{k}}^S,
\end{equation}
where in the first term, $E_{c\mathbf{k}}-E_{v\mathbf{k}}$ is the energy difference between the conduction ($c$) and valence ($v$) Kohn-Sham states corrected within the GW approximation, with crystal momentum $\mathbf{k}$. The second term contains the interaction between electron-hole pairs, with $K_{vc\mathbf{k},v'c'\mathbf{k}'}(\Omega^S)$, the energy dependent electron-hole interaction kernel. When solved, we obtain exciton eigenenergies, $\Omega^S$, where $S$ is the principal quantum number, and $A_{vc\mathbf{k}}^S$ are the exciton expansion coefficients. The latter can be used to construct a real-space representation of the exciton, $\Psi_{S}(\mathbf{r}_e,\mathbf{r}_h) = \sum_{cv\mathbf{k}} A_{vc\mathbf{k}}^S \psi_{c\mathbf{k}}(\mathbf{r}_e)\psi^\star_{v\mathbf{k}}(\mathbf{r}_h)$, where $\psi_{n\mathbf{k}}(\mathbf{r})$ denote single-particle Kohn-Sham Bloch eigenstates.
For 2D TMDs, the kernel is dominated by the attractive direct term
\begin{equation}
\label{eq6}
    K_{vc\mathbf{k},v'c'\mathbf{k'}}^d(\Omega^S) = - \sum_{\mathbf{GG'}} M_{c'c}^*(\mathbf{k},\mathbf{q},\mathbf{G})M_{v'v}(\mathbf{k},\mathbf{q},\mathbf{G'}) \tilde{\varepsilon}^{-1}_{\mathbf{GG}'; cvc'v'\mathbf{k}}(\mathbf{q},\Omega^S) v(\mathbf{q}+\mathbf{G'}),
\end{equation}
where $M_{n'n}(\mathbf{k},\mathbf{q},\mathbf{G})=\braket{n'\mathbf{k}+\mathbf{q}|e^{i(\mathbf{q}+\mathbf{G})\cdot\mathbf{r}}|n\mathbf{k}}$ are the plane-wave matrix elements, and $\tilde{\varepsilon}_{\mathbf{GG}'}^{-1}$ is the effective inverse dielectric function that contains the important dynamical and local-field effects~\cite{strinati1982,strinati1984,strinati1988,spataru2010,spataru2013}:
\begin{multline}
\label{eq7}
\tilde{\varepsilon}^{-1}_{\mathbf{GG}',cvc'v'\mathbf{k}}(\mathbf{q},\Omega^S) = \epsilon_{\mathbf{GG}'}^{-1}(\mathbf{q},0) - \frac{1}{\pi} P \int_0^{\infty} d\omega~ \text{Im}{\epsilon_{\mathbf{GG}'}^{-1}(\mathbf{q},\omega)} \\
    \times \left[ \frac{2}{\omega} + \frac{1}{\Omega^S - \omega - (E_{c\mathbf{k}+\mathbf{q}}-E_{v'\mathbf{k}})} + \frac{1}{\Omega^S - \omega - ( E_{c'\mathbf{k}}-E_{v\mathbf{k}+\mathbf{q}})} \right].
\end{multline}
For intrinsic semiconductors, $\Omega^S - (E_c-E_v)$ is small in comparison to the plasmon energy and the second term on the right-hand side of Eq.~\ref{eq7} can be neglected~\cite{rohlfing2000}; this is the so-called static approximation to the BSE. For example, in the case of organic crystals for which the difference between exciton binding energy and plasma frequency is about a factor of ten, Zhang et al. showed that corrections due to dynamical screening are about 15\% of the exciton binding energy.~\cite{zhang2023} For doped semiconductors, when the carrier plasmon energies are below or nearly resonant with the exciton binding energy, the plasmon and exciton are dynamically coupled and the static approximation fails dramatically~\cite{lischner2013,zhou2015,gao2016}. An accurate dynamical treatment~\cite{spataru2010,spataru2013,ando1982}, which both retains local-field effects and elucidates the role of exciton eigenvalue self-consistency, is required to predict and interpret optical spectra of low-dimensional systems at experimentally relevant doping concentrations.\

\begin{figure*}[h!]
	\includegraphics[width=0.85\textwidth]{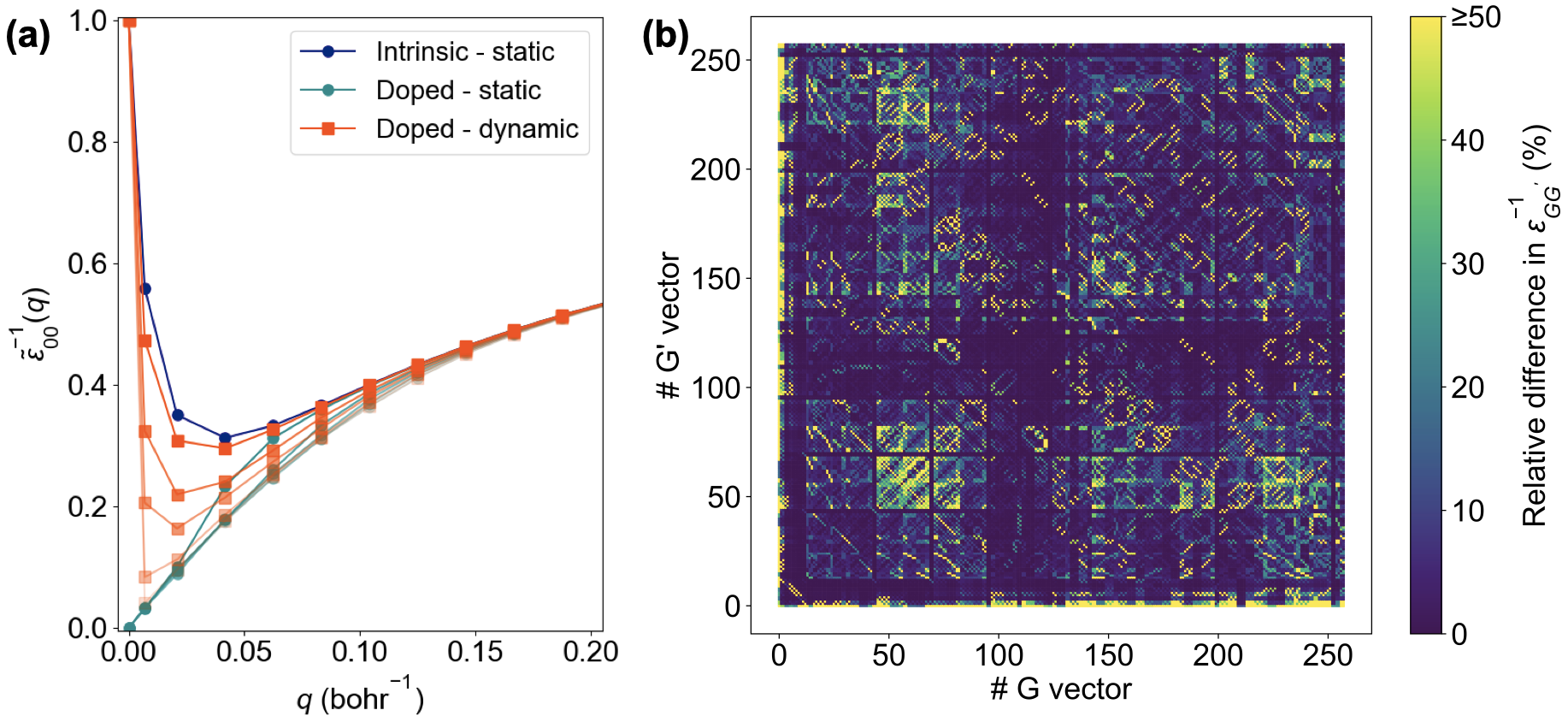}
	\centering
	\caption{(a) Head element of the static inverse dielectric matrix, $\epsilon_{00}^{-1}(\mathbf{q},0)$, plotted for small $\mathbf{q}$ wavevectors for the intrinsic (blue curve), and the static doped MoTe$_2$ monolayer systems (cyan curves). Orange curves correspond to the effective screened dielectric function, $\tilde{\varepsilon}_{00}^{-1}(\mathbf{q},E_b^{S})$, defined in Eq.~\ref{eq8} for doped monolayers. Each cyan and orange curve corresponds to a given doping concentration; the curve transparency increases with the doping density. (b) Relative difference $( \epsilon_{\text{int},\mathbf{GG}'}^{-1} - \epsilon_{\mathbf{GG'}}^{-1})/\epsilon_{\text{int},\mathbf{GG}'}^{-1}$ for all $\mathbf{G}$ and $\mathbf{G'}$ combinations at $q=0.02$\,bohr$^{-1}$ and $\omega=0$. Data are shown for a carrier density of $1.6 \times 10^{12}$\,cm$^{-1}$. The full set of data for various carrier concentrations can be found in Figure~S3.}
	\label{fig2}
\end{figure*}

Here, we develop a PPM for the real part of the dielectric function with the form
\begin{equation}
\label{eq2}
    \text{Re}~\epsilon^{-1}_{\mathbf{GG}'}(\mathbf{q},\omega) = \epsilon_{\mathrm{int},\mathbf{GG}'}^{-1}(\mathbf{q},\omega) + \frac{\Omega^2_{\mathbf{GG}'}(\mathbf{q})}{\omega^2-\tilde{\omega}_{\mathbf{GG}'}^2(\mathbf{q})},
\end{equation}
where $\epsilon_{\mathrm{int},\mathbf{GG}'}^{-1}(\mathbf{q},\omega)$ is the real part of inverse dielectric matrix computed for the intrinsic system using a plasmon-pole model as described in the SI, $\Omega_{\mathbf{GG}'}^2(\mathbf{q})$ is the plasmon-pole strength, and $\tilde{\omega}_{\mathbf{GG}'}(\mathbf{q})$ corresponds to the carrier plasmon frequency, which depends on both the wavevector $\mathbf{q}$ and the reciprocal lattice vectors $\mathbf{G}$ and $\mathbf{G}'$. In order to parameterize $\Omega^2_{\mathbf{GG}'}(\mathbf{q})$ and $\tilde{\omega}_{\mathbf{GG}'}^2(\mathbf{q})$, we constrain the PPM to exactly recover the computed value of $\text{Re}~\epsilon^{-1}_{\mathbf{GG}'}(\mathbf{q},\omega)$ for two real frequencies $\omega = 0$ and $\omega = \omega_0$, in analogy with the Godby-Needs approach~\cite{godby1989,oschlies1995} but adapted here for doped systems. We note here that the choice of the reference frequency $\omega_0$ is arbitrary, as verified in Figure~S1. The plasmon pole strength is given by
\begin{equation}
\label{eq3}
    \Omega_{\mathbf{GG}'}^2(\mathbf{q}) = \omega_0^2 \frac{[\epsilon_{\mathrm{int},\mathbf{GG}'}^{-1}(\mathbf{q},0)-\epsilon_{\mathbf{GG}'}^{-1}(\mathbf{q},0)][\epsilon_{\mathrm{int},\mathbf{GG}'}^{-1}(\mathbf{q},0)-\epsilon_{\mathbf{GG}'}^{-1}(\mathbf{q},\omega_0)]}{[\epsilon_{\mathbf{GG}'}^{-1}(\mathbf{q},0)-\epsilon_{\mathbf{GG}'}^{-1}(\mathbf{q},\omega_0)]},
\end{equation}
and the plasmon frequency by
\begin{equation}
\label{eq4}
    \tilde{\omega}_{\mathbf{GG}'}^2(\mathbf{q}) = \omega_0^2 \frac{[\epsilon_{\mathrm{int},\mathbf{GG}'}^{-1}(\mathbf{q},0)-\epsilon_{\mathbf{GG}'}^{-1}(\mathbf{q},\omega_0)]}{[\epsilon_{\mathbf{GG}'}^{-1}(\mathbf{q},0)-\epsilon_{\mathbf{GG}'}^{-1}(\mathbf{q},\omega_0)]}.
\end{equation}
In principle, the off-diagonal components of the dielectric function of systems without inversion symmetry have non-zero imaginary components, as indeed observed for both the intrinsic and doped MoTe$_2$ monolayers (Figure~S2a). Including the imaginary components in our PPM is however not trivial~\cite{zhang1989}. However, as discussed in Ref.~\citenum{qiu2016} for 2D materials, most off-diagonal matrix elements ($\mathbf{G}\ne \mathbf{G}'$) have non-zero $G_z$ and $G_{z’}$ components and integrating out the screening in the out-of-plane direction zeroes out the imaginary components of the dielectric matrix. This is verified by computing the effective 2D inverse dielectric function, $\epsilon_{\mathrm{2D}}^{-1}$, defined in Ref.~\citenum{qiu2016}, shown in Figure~S2b. Additional details are reported in the SI. In Figure~\ref{fig1}b, we show the frequency dependence of the difference in the effective 2D inverse dielectric matrix, $\delta \epsilon_{\mathrm{2D}}^{-1}(\mathbf{q},\omega) = \epsilon_{\mathrm{2D}}^{-1}(\mathbf{q},\omega) - \epsilon_{\mathrm{int},\mathrm{2D}}^{-1}(\mathbf{q},\omega)$ for a series of small wavevectors $\mathbf{q}$, for monolayer MoTe$_2$. Our PPM reproduces the frequency dependence of the effective 2D dielectric function obtained from a FF calculation for the considered series of wavevectors. With increasing $\mathbf{q}$, the carrier plasmon-pole shifts to higher frequency, while its amplitude decreases and becomes negligible for $|\mathbf{q}|>0.20$\,bohr$^{-1}$. This effect is further highlighted in Figure~\ref{fig2}a, where the head matrix elements ($\mathbf{G}=\mathbf{G}'=0$) of $\epsilon^{-1}(\mathbf{q})$ are represented for small wavevectors $\mathbf{q}$. For the intrinsic MoTe$_2$ monolayer, the static head matrix element goes to unity as $\mathbf{q}\to 0$, which reflects the absence of screening at long wavelengths in 2D semiconductors where the only extrinsic dielectric medium is vacuum~\cite{deslippe2012,ismail2006}. For the doped systems, the static head matrix element goes to zero as $\mathbf{q}\to 0$, reflecting the larger screening induced by the presence of the charged free carriers.

While previous GW-BSE calculations of carrier-doped TMD monolayers assumed that only the head matrix element was strongly modified with doping~\cite{liang2015,gao2016}, we find significant differences between the matrix elements of a doped and intrinsic MoTe$_2$ monolayer for several $\mathbf{G}$ and $\mathbf{G'}$ combinations. In Figure~\ref{fig2}b, we show, for all $\mathbf{G}$ and $\mathbf{G'}$ combinations, the relative variation in the static matrix elements $( \epsilon_{\text{int},\mathbf{GG}'}^{-1} - \epsilon_{\mathbf{GG}'}^{-1})/\epsilon_{\text{int},\mathbf{GG}'}^{-1}$, for a wavevector $q=0.02$\,bohr$^{-1}$. While the largest absolute difference is indeed found for the head matrix element ($\mathbf{G} = \mathbf{G}' = 0$), we find non-negligible contributions from carrier screening to all matrix elements, ultimately giving rise to local-field effects and highlighting the importance of going beyond the spatially uniform screening background approximation when modeling the carrier screening in doped low-dimensional systems (Figure~\ref{fig1}c$-$d).\

Next, we use our new PPM to obtain the frequency-dependent dielectric matrix of a doped monolayer MoTe$_2$ for six different doping densities. The inverse dielectric matrix is then used to compute the screened Coulomb interaction (Eq.~\ref{eq1}), used itself to obtain the QP energies, within the GW approximation.\

Solving the BSE with the fully frequency-dependent dielectric matrix with the form of Eq.~\ref{eq7} is impractical, and a PPM can greatly reduce computational complexity without loss of accuracy. Assuming that the carrier plasmon-pole dominates the low-frequency screening and that the exciton is well-localized in $k$-space, such that $\sum_{cv\mathbf{k}} | A^S_{cv\mathbf{k}} |^2(E_{c\mathbf{k}}-E_{v\mathbf{k}}) - \Omega_S = E_{\mathrm{int}}^S$, where $E_{\mathrm{int}}^S$ can be approximated as the exciton binding energy $E_b^S = E_g^{\mathrm{QP}}-\Omega^S$, Eq.~\ref{eq7} can be simplified to
\begin{equation}
\label{eq8}
    \tilde{\varepsilon}_{\mathbf{GG}'}^{-1} \left(\mathbf{q}, E^S_b \right) = \epsilon_{\mathbf{GG}'}^{-1}(\mathbf{q},0) + \delta \tilde{\varepsilon}^{-1}_{\mathbf{GG}'}(\mathbf{q}, E^S_b)
\end{equation}
with
\begin{equation}
\label{eq9}
    \delta \tilde{\varepsilon}_{\mathbf{GG}'}^{-1} \left(\mathbf{q}, E^S_b \right) =  \frac{\Omega_{\mathbf{GG}'}^2(\mathbf{q})}{\tilde{\omega}_{\mathbf{GG}'}^2(\mathbf{q})} \frac{E_b^S}{\tilde{\omega}_{\mathbf{GG}'}(\mathbf{q}) + E_b^S},
\end{equation}
where $\Omega_{\mathbf{GG}'}^2$ and $\tilde{\omega}_{\mathbf{GG}'}^2$ are defined in Eqs.~\ref{eq3} and \ref{eq4}, respectively, and the exciton binding energy $E_b^S$ is self-consistently computed. This expression only requires calculation of the intrinsic and doped static dielectric matrix, a considerable simplification and storage memory reduction relative to treating the full-frequency dielectric matrix.\

The head ($\mathbf{G}=\mathbf{G}'=0$) of the inverse effective dielectric matrix for doped monolayer MoTe$_2$ evaluated at $\hbar \omega = E_b$ is depicted in Figure~\ref{fig2}a, revealing noticeable deviations from the static case, with a sharp increase to unity as $\mathbf{q} \to 0$. Such behavior is accurately described only when the effective inverse dielectric function is computed self-consistently, as shown in Figure~S4. This sharp increase to unity is consistent with the carrier plasmon pole frequency tending to zero for very long wavelength ($\mathbf{q}\to0$), hence removing any screening contribution from the charged free carriers. \

\begin{figure*}[h!]
	\includegraphics[width=0.98\textwidth]{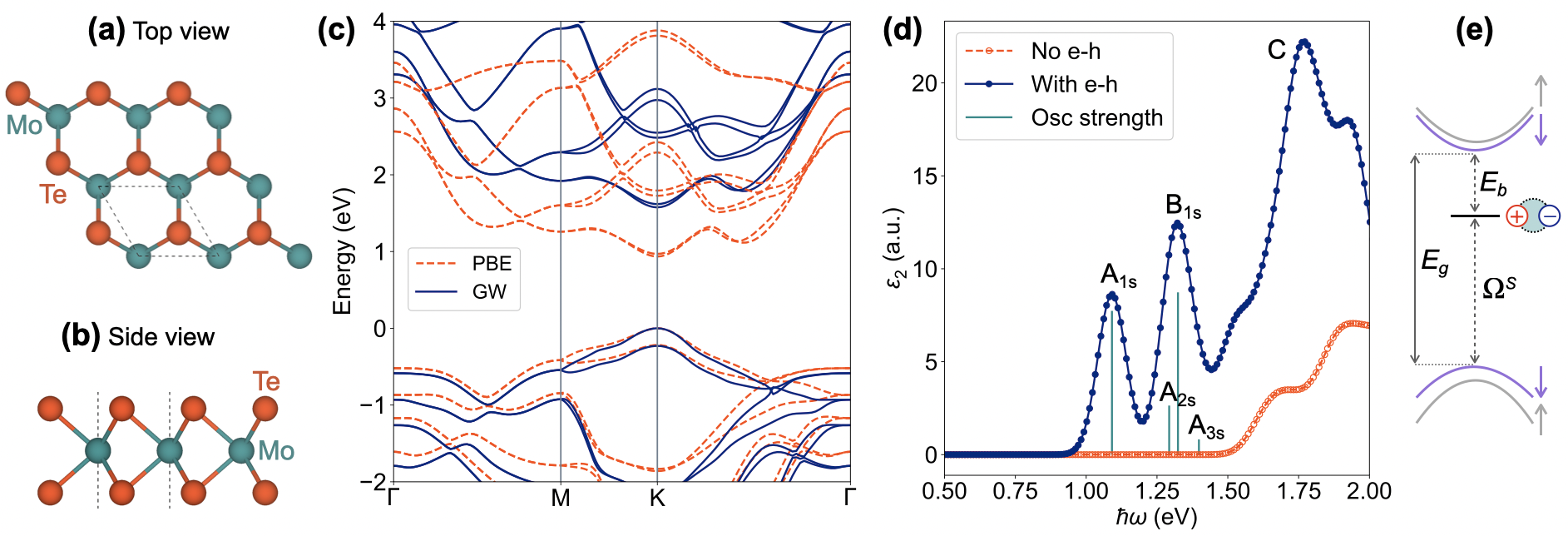}
	\centering
	\caption{(a) Top and (b) side views of monolayer 2H MoTe$_2$. (c) Monolayer MoTe$_2$ electronic band structure computed with DFT-PBE (dashed orange line) and GW (solid blue line). (d) Absorption spectrum of monolayer MoTe$_2$ without (orange curve) and with (blue curve) electron-hole interactions, using a constant broadening of 50\,meV. $A$ and $B$ exciton energies are represented with vertical cyan lines, with spin-orbit splitting of the VBM ($230$\,meV) and CBM ($43$\,meV). (e) Representation of the fundamental gap $E_g$, the exciton energy $\Omega^S$ and the corresponding binding energy $E_b = E_g - \Omega^S$ for the first electron-hole bound state.}
	\label{fig3}
\end{figure*}

Next, the effective dynamical dielectric matrix defined in Eq.~\ref{eq8}$-$\ref{eq9} is used to solve the BSE and compute the excited state properties of a doped monolayer MoTe$_2$. There is a limited number of works that have investigated the electronic and optical properties of monolayer 2H MoTe$_2$~\cite{ramasubramaniam2012,ruppert2014,lezama2015,yang2015,koirala2016,robert2016}, whose structure is shown in Figure~\ref{fig3}a$-$b. Moreover, a spread of QP gap values are reported in prior computational studies, which reflect either an insufficient level of theory or a lack of convergence, as pointed our in Ref.~\cite{qiu2016}, further motivating our calculations. 

Here, we compute the QP band structure within the GW approximation (Figure~\ref{fig3}c), and report a direct gap at $K$ of $1.58$\,eV ($1.71$\,eV) with (without) spin-orbit coupling (SOC). Compared with other Mo-based monolayer TMDs, MoTe$_2$ has a stronger SOC which splits the valence band maximum (VBM) by $230$\,meV and the conduction band minimum (CBM) by $43$\,meV, at the $K$ point of the Brillouin zone~\cite{robert2016}.\

The presence of strongly bound excitons in MoTe$_2$ explains the large difference between the QP gap and the measured optical gap of $1.1$\,eV~\cite{biswas2023,ruppert2014}. Indeed, for 2D TMDs, the photoluminescence spectrum is dominated by exciton transitions, rather than band-to-band recombination (Figure~\ref{fig3}e)~\cite{qiu2013}. The imaginary part of the frequency-dependent dielectric function $\epsilon_2(\omega)$, obtained from a GW-BSE calculation assuming incident light polarized in the plane of the monolayer, is displayed in Figure~\ref{fig3}d. As expected, the inclusion of electron-hole interactions leads to good agreement with the experimental optical gap of $1.1$\,eV~\cite{biswas2023,ruppert2014}. The spin-orbit splitting of the VBM and CBM, depicted in Figure~\ref{fig3}e, leads to two distinct exciton series, with the $A$ and $B$ exciton peak, respectively, at $1.09$\,eV and $1.35$\,eV. Transitions associated with the $2s$ and $3s$ states of the $A$ excitons are included under this second peak, and fall at an energy of $1.31$ and $1.43$\,eV, respectively. Relevant results for the exciton states are reported in Table~\ref{tab2} and are in very good agreement with energies obtained from our photoluminescence measurements, which are detailed elsewhere (Ref.~\citenum{biswas2023}). The slight remaining discrepancy is likely due to the dielectric environment, \textit{i.e.}, hBN substrate, present in the experimental setup and not considered in the calculations. Computational parameters are reported in the SI.\

\begin{table}[h!]
\centering
\caption{Exciton energy (in eV), exciton binding energy (in meV), and squared dipole moment, Dip. Mom. (in arbitrary units), of $A$ and $B$ excited states, computed with GW-BSE, and compared to our experimental PL data detailed in Ref.~\citenum{biswas2023}. \label{tab2}}
\begin{tabular}{c|cccc}
\hline
\hline
Excited & \multicolumn{3}{c}{BSE} & EXP~\cite{biswas2023}\\
states & $\Omega$ (eV) & $E_b$ (meV) & Dip. Mom. (a.u.) & $\Omega$ (eV) \\
\hline
$A_{1s}$ & $1.09$ & $490$ & $0.22 \times 10^5$ & $1.17$ \\ 
$A_{2s}$ & $1.31$ & $270$ & $0.20 \times 10^4$ & $1.29$ \\
$B_{1s}$ & $1.35$ & $230$ & $0.19 \times 10^5$ & $1.40$ \\
$A_{3s}$ & $1.43$ & $150$ & $0.12 \times 10^4$ & $1.33$ \\
\hline
\hline 
\end{tabular}
\end{table}

\begin{figure*}[h!]
	\includegraphics[width=0.98\textwidth]{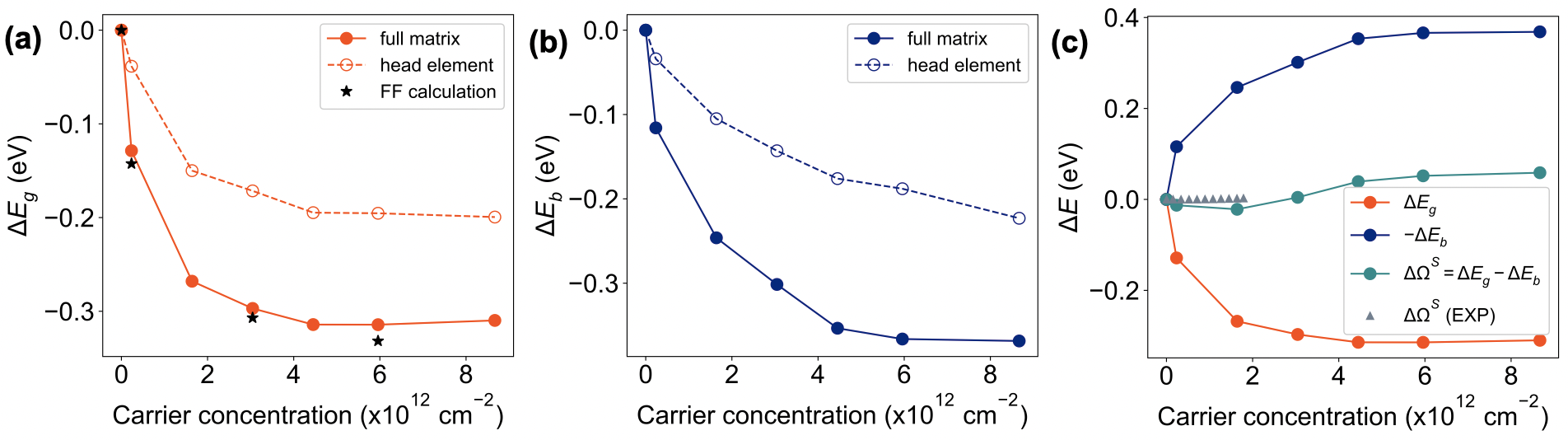}
	\centering
	\caption{For monolayer MoTe$_2$, doping dependence of (a) the QP band gap obtained with our PPM (Eq.~\ref{eq2}) with the carrier plasmon screening included in the head matrix element only ($\delta \epsilon_{00}^{-1}(q, \omega)$ in the right-hand side of Eq.~\ref{eq2}) or in the full dielectric matrix ($\delta \epsilon_{\mathbf{G}\mathbf{G}'}^{-1}(q, \omega)$ in Eq.~\ref{eq2}), compared to a full-frequency calculation (black star markers), (b) the exciton binding energy obtained self-consistently from BSE calculations with the effective dielectric function accounting for dynamical effects in the head matrix element $\delta \tilde{\varepsilon}_{00}^{-1}(\mathbf{q},E_b^S)$ only or in the full matrix $\delta \tilde{\varepsilon}_{\mathbf{GG}'}^{-1}(\mathbf{q},E_b^S)$ (Eq.~\ref{eq8}$-$\ref{eq9}), and (c) the computed exciton energy with the PPM (cyan curve) obtained as $\Delta \Omega = \Delta E_g - \Delta E_b$, compared to experimental data (gray markers) from Ref.~\citenum{biswas2023}.}
	\label{fig4}
\end{figure*}

The presence of charged free carriers significantly affects both the electronic and optical properties of 2D semiconductors, as it strongly modifies the electric background of the monolayer. As a result of two competing effects, a weakening of the electron-electron interaction and a decrease in the carrier-occupation energy~\cite{liang2015}, we observe a large renormalization of the QP gap with increasing doping density (Figure~\ref{fig4}a). At relatively low carrier density, the gap drops sharply from $1.71$\,eV down to $1.44$\,eV at a carrier density of $1.6 \times 10^{12}$\,cm$^{-2}$, eventually reaching $1.40$\,eV at a higher carrier density of $8.7 \times 10^{12}$\,cm$^{-2}$. At even higher doping concentration, the QP gap is expected to increase, due to the Burstein-Moss shift of the continuum energy~\cite{moss1954,burstein1954} as free carriers populate the conduction band. Our PPM works remarkably well to predict the band gap renormalization, with a maximal deviation of 17\,meV with respect to a FF calculation performed within the static subspace approximation. Figure~\ref{fig4}a also highlights the underestimation in the screening when assuming that only the head of the dielectric function is modified by the presence of free carriers; the use of a single plasmon pole~\cite{gao2016} overestimates the screening and, hence, the QP gap renormalization (Figure~S5a).

\begin{figure*}[h]
	\includegraphics[width=0.98\textwidth]{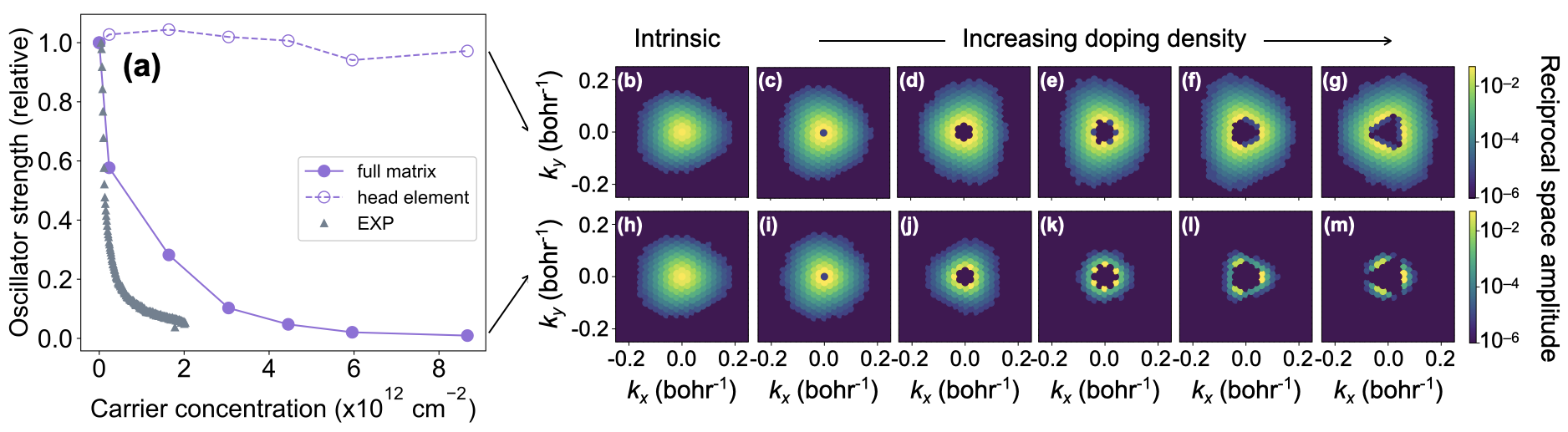}
	\centering
	\caption{(a) Evolution of the oscillator strength with the doping density for monolayer MoTe$_2$, obtained from a self-consistent BSE calculation using the effective dielectric function defined in Eq.~\ref{eq8}$-$\ref{eq9}, with carrier plasmon screening accounted for in all matrix elements, $\delta \tilde{\varepsilon}_{\mathbf{GG}'}^{-1}(\mathbf{q},E_b^S)$, or in only the head matrix element, $\delta \tilde{\varepsilon}_{00}^{-1}(\mathbf{q},E_b^S)$. Experimental data (gray markers) from Ref.~\citenum{biswas2023} are shown for comparison. (b)$-$(m) Representation of the lowest-energy exciton envelope wavefunction in $k$-space (normalized such that $\sum_{\mathbf{k}} |A(\mathbf{k})|^2 = 1$), centered around K$(0,0)$, for the intrinsic MoTe$_2$ monolayer, and doped monolayer MoTe$_2$, with doping density of $2.3 \times 10^{11}$, $1.6 \times 10^{12}$, $3.0 \times 10^{12}$, $4.5 \times 10^{12}$, and $5.9\times 10^{12}$\,cm$^{-2}$. Plots (b)$-$(g) are obtained with dynamical effects only in the head matrix element, and (h)$-$(m) with dynamical effects in all matrix elements. Two effects are visible on these plots: a localization of the exciton in $k$-space with increasing doping, corresponding to a delocalization in real space, and Pauli blocking effect that blocks some vertical transitions at $K$ and around.}
	\label{fig5}
\end{figure*}

Excitonic and optical properties, including exciton binding energy, exciton energy, and oscillator strength, are also altered by the presence of charged free carriers, as shown in Figures~\ref{fig4} and \ref{fig5}. As expected from the carrier-induced screening of the electron-hole Coulomb interaction combined with Pauli blocking effects that prevent direct vertical transitions at the $K$ valley (Figure~\ref{fig5}), the exciton binding energy decreases with doping, by as much as $250$\,meV at a doping density of $1.6 \times 10^{12}$\,cm$^{-2}$, and eventually saturates at higher carrier concentration. Figure~\ref{fig4}b also shows that restricting inclusion of dynamical effects from carrier plasmon screening to the head matrix element (instead of the full matrix) leads to a much smaller reduction in exciton binding energy, highlighting the importance of off-diagonal matrix elements and, hence, local-field effects. In contrast, a static calculation clearly overestimates the screening for free carriers and so does a single plasmon-pole approach~\cite{gao2016}, as reported in Figure~S5b. Notably, omitting to self-consistently (SC) compute the exciton binding energy in Eq.~\ref{eq9} underestimates both the screening and reduction in exciton binding energy with increasing doping (Figure~S4 and S5b).

The reduction in exciton binding energy almost exactly compensates for the renormalization of the QP gap, resulting in an almost constant exciton energy with doping density. This is shown in Figure~\ref{fig4}c, which also validates the efficiency of our PPM, as the theoretical results are in good qualitative and quantitative agreement with photoluminescence data reported in Ref.~\citenum{biswas2023}, in the experimentally-accessible doping region. The corresponding computed absorption spectra can be found in Figure~S7. A similar conclusion for a doped MoS$_2$ monolayer was reported in a prior work~\cite{gao2016}, with a more approximate treatment of the dynamical screening associated with the free carriers, as emphasized in Figure~S4c. Here, we go beyond these approximations and obtain good agreement with experiments for the exciton energy, and quantitative predictions for QP band gap, as verified with a FF calculation.\

Figure~\ref{fig5}a shows the change in oscillator strength for the first exciton, relative to the intrinsic system, compared to experimental data obtained for a carrier-doped MoTe$_2$ monolayer~\cite{biswas2023}. A decrease in oscillator strength with increasing doping density is observed and can be explained as originating from two effects: Pauli blocking and delocalization of the exciton in real space, due to a weakening of the electron-hole interaction from carrier screening. The latter is further verified by plotting the exciton wavefunction, shown at various doping densities in Figure~\ref{fig5}h$-$m; the exciton becomes more localized in $k$-space, corresponding to a delocalization in real-space. Since optical absorption occurs when the electron and hole positions are identical, the probability for absorption is reduced as the exciton delocalizes, explaining the reduced oscillator strength. The importance of solving the BSE self-consistently for the doped systems is evidenced from Figure~S6. Moreover, while neglecting local-field effects leads to an incorrect doping-dependence of the oscillator strength and exciton wavefunction (Figure~\ref{fig5}b$-$g), our modified PPM shows remarkable agreement with photoluminescence peak intensity at low doping density.\

In conclusion, we have developed a new, computationally-efficient approach to predict the electronic and optical properties of doped semiconductors within the \textit{ab initio} GW-BSE framework, and we have applied it to monolayer MoTe$_2$, a 2D material of significant interest and compared it to detailed spectroscopy experiments~\cite{biswas2023}. We find that as a function of carrier density, the band gap renormalization and the reduction of the exciton binding energy have a similar doping dependence, resulting in an almost constant exciton energy for the lowest excited state. This is in agreement with recent photoluminescence measurements where the onset of the optical spectrum remains unchanged with increased doping density. This agreement is made possible by careful treatment of the dynamical effects in the effective screened electron-hole interaction, accounting for the off-diagonal terms responsible for local-field effects. As our new PPM can further be applied to treat the charge carrier screening in any 2D material with delocalized excitons or van der Waals homobilayer, we expect it will deepen our understanding of the underlying physics in realistic intrinsically doped 2D devices.

\begin{suppinfo}

The Supporting Information is available free of charge.
\begin{itemize}
  \item \textit{Ab initio} calculations parameters; effective 2D dielectric function; parametrization, performance, and efficiency of the PPM; absorption spectra for doped MoTe$_2$ monolayers (PDF).
\end{itemize}

\end{suppinfo}

\begin{acknowledgement}

The calculations in this work were primarily supported by the Center for Computational Study of Excited-state Phenomena in Energy Materials (C2SEPEM), funded by the US Department of Energy (DOE) under contracts No.~DE-AC02-05CH11231 and DE-FG02-07ER46405. The Theory of Materials FWP at LBNL, funded by the DOE under contract No.~DE-AC02-05CH11231, supported the development of the plasmon-pole model. Computational resources are provided by the National Energy Research Scientific Computing Center (NERSC). A.C acknowledges the support from Wallonie Bruxelles International under contract No.~SUB/2021/512815.

\end{acknowledgement}

\bibliography{references.bib}

\begin{figure*}[h]
	\includegraphics[width=0.5\textwidth]{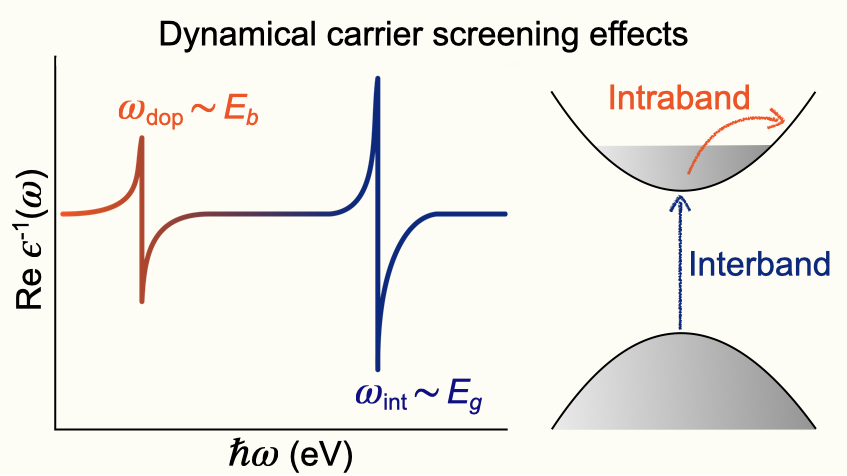}
	\centering
	\caption{For Table of Contents Only}
	\label{toc}
\end{figure*}

\end{document}